\documentstyle[12pt]{article}
\newcommand{\refjl}[4]{{#1 }{\em #2 }{\bf #3 }{#4}}
\newcommand{\refbk}[3]{{#1 }{\em #2 }{#3}}

\def\bmax{b_{\rm max}}
\def\etal{{\em et al.\hspace{1ex}}}
\def\2D{{\small \bf 2{\sc D}}\hspace{1ex}}

\def\JPA{{J. Phys. A:Math. Gen.}}
\def\JPC{{J. Phys. C:Solid State Phys.}}
\def\PRL{{Phys. Rev. Lett.}}
\def\PRB{{Phys. Rev. B}}
\def\dash{{------\hspace{1ex}}}

\def\figpivot{Fig.~1} 
\def\figbetavnt{Fig.~2} 
\def\figsusbht{Fig.~3} 
\def\figmagucamp{Fig.~4} 
\def\figsusucamp{Fig.~5} 
\def\figbetagrt{Fig.~6}

\def\tablespin1ser{Table~I}
\def\tablespinhalfser{Table~II}
\def\tablegammauc{Table~III}
\def\tablealphauc{Table~IV}
\def\tableampl{Table~V}
\def\tablebetadelta{Table~VI}
\def\tablegammadelta{Table~VII}

\def\tabcorr{Table~A-I}

\begin{document}
\setcounter{page}{0}
\thispagestyle{empty}

\title{Low-Temperature Series Expansions for the Spin-1
Ising Model.}

\author{
I. G. Enting\thanks{CSIRO, Division of Atmospheric Research,
Private Bag 1, Mordialloc, Vic. Australia 3195.\newline 
\mbox{\hspace{4ex} e-mail: ige@dar.csiro.au}}, 
A. J. Guttmann\thanks{Department of Mathematics, 
The University of Melbourne, Parkville, Vic. Australia 3052.\newline 
\mbox{\hspace{4ex} e-mail: tonyg, iwan@maths.mu.oz.au}}\hspace{1ex}
and I. Jensen$^{\dagger}$}

\maketitle

\thispagestyle{empty}

\begin{abstract}
The finite lattice method of series expansion has been used
to extend low-temperature series for the partition function,
order parameter and susceptibility of the spin-1 Ising model on
the square lattice. A new formalism is described that uses two 
distinct transfer matrix approaches in order to significantly reduce 
computer memory requirements and which permits the derivation 
of the series to 79th order. Subsequent analysis of the series 
clearly confirms that the spin-1 model has the same dominant 
critical exponents as the spin-$\frac{1}{2}$ Ising model. Accurate 
estimates for both the critical temperature and non-physical 
singularities are obtained. In addition, evidence for a non-analytic 
confluent correction with exponent $\Delta_{1} = 1.1 \pm 0.1$ is found.
\end{abstract}

\newpage     

\section{Introduction}

Low temperature expansions for the spin-1 Ising model were first 
obtained by Fox and Guttmann (1973), who gave a 26 term series
for the square lattice, of which only the first 24 terms were 
correct. The method used to obtain the series was a generalisation
of the code method of Sykes \etal\ (1965). Series on other lattices,
both two- and three-dimensional were also obtained. Subsequently 
the finite lattice method of series expansions
(de Neef, 1975; de Neef and Enting, 1977) has proved to be an
extremely powerful technique for deriving series expansions
for a range of two-dimensional models. The formalism
is applicable in higher dimensions but the technique becomes
progressively less efficient (Guttmann and Enting, 1993).
Adler and Enting (1984) used the finite lattice method
to extend low temperature expansions for the zero-field
partition function, the magnetisation and the zero-field
susceptibility of the spin-1 Ising model to order $u^{45}$.
As we have noted elsewhere, developments in computing over the last
decade, notably faster computers with more memory, have allowed 
larger finite lattice calculations to be made. By re-running the 
program used by  Adler and Enting we easily extended the series 
to 65 terms. We have, however, recently implemented a revised 
algorithm that removes much of the memory-size requirement that 
has previously limited our finite-lattice calculations.
This involves using two different ways of calculating finite lattice
partition functions. A preliminary analysis of the formalism
was given by Enting (1990). In this paper we use this new formalism 
to calculate low-temperature spin-1 Ising series to order $u^{78}$.
Using a rather cumbersome correction method described in the appendix
we also obtained the coefficients of $u^{79}$. We have also 
substantially extended the spin-$\frac{1}{2}$ low-temperature
susceptibility series.

The layout of the paper is as follows: In Section 2 we describe
the finite lattice method of series expansions. The various
expansions are detailed in Section 3. The results of the
series analysis, with the emphasis on the new extended spin-1
series, are presented in Section 4. Finally, Section 5 contains
a short summary and discussion of our results.

\section{Series expansions from the finite lattice method}

As in the study by Adler and Enting (1984), the series
expansions are derived from
$$
Z \approx \prod_{m,n} Z_{mn}^{ a_{mn}}
\qquad\mbox{with $m \leq n$ and $m+n \leq k$}
\eqno(2.1)
$$
where $Z$ is the infinite lattice partition function and the $Z_{mn}$
are the partition functions of the $m \times n$ lattices. The
weights, $a_{mn}$ are derived from the expressions given by
Enting (1978), modified to exploit the rotational symmetry
of the lattice. The difference from Enting and Adler is that
we use a larger cut-off, $k$, which leads to longer series.

The finite lattice method relies on
efficient techniques for evaluating the $Z_{mn}$.
We use what are known as `transfer matrix'
techniques. These work by moving a boundary through the lattice
and constructing a partial sum of Boltzmann
weights for each possible configuration of the boundary.
The traditional form of transfer matrix calculation involves
moving the boundary one column at a time.
For a system with $q$ states per site, evaluating
$Z_{mn}$ involves $n$ iterations of $q^{2m}$ operations
on series. It is more efficient to move the boundary
by adding one site at a time. Evaluating $Z_{mn}$ involves 
$m\times n$ iterations of $q^{m+1}$ series operations.
The `one-site-at-a-time' algorithm seems to have
been rediscovered independently a number of times. Our use of the 
technique derives from unpublished work by R.J. Baxter.

The new procedure  proposed by Enting (1990)
and adopted here is to use  (2.1) as before
but to use two different techniques for calculating the $Z_{mn}$.
We define a cut-off parameter $\bmax$ so that, for a $q$-state system,
the maximum vector size is $q^{\bmax}$. In evaluating $Z_{mn}$
(and considering only $m \leq n$ because of our use of symmetry),
 if $m \leq \bmax$ we
use our original procedure of building up the lattice column
by column with each column built up one site at a time.
Evaluating $Z_{mn}$ requires $mn q^{m+1}$ series operations
but the evaluation of $Z_{mn}$ enables us to
determine $Z_{mp}$ for $p<n$ with little extra computation.
For square (or nearly square) lattices we evaluate the
partition functions by a technique in which the boundary
pivots about a central point. The general principle is based
on unpublished work by R.J. Baxter.

The `pivoting' transfer matrix approach use 3 integers, $a$, $b$
and $c$ (with $c=b$ or $c = b-1$) to specify the rectangles.
The rectangles are of size $m = (a+b)$ by $n = (b+c+1)$. We refer 
to sites by integer coordinates $(x,y)$ with $1-a \leq x \leq b$ 
and  $-c \leq y \leq b$. The partition function is a sum over the 
$q^a$ conditional lattice sums in which the $a$ sites $(x,0)$ with 
$x \leq 0$ are fixed. Transfer matrix techniques are used
to calculate the lattice sum conditional on the state of the
fixed sites.

The algorithm outlined below requires space for $2q^b$ series and
takes time $\propto m\times n \times q^{(a+b)}$.

We consider two alternative forms of `cut-off'

{\it Space-limited:} If the execution time was not limiting then 
the smallest rectangle that could not be computed by pivoting
would be the square of size $(2b+2)\times (2b+2)$. Thus we use
$$
m+n \leq k = 4\bmax+2
$$
We use the original transfer matrix technique for $m \leq \bmax$
and use the pivoting algorithm for rectangles of size $m =(b+c+1)$ 
by $n =(a+b)$ for $b$, $c \leq \bmax$. For $m=\bmax+1$ the 
longest rectangle is of length $n= a+b = 3\bmax+1$, so that in terms 
of the cut-off, $k$ the time requirement grows as $q^{3k/4}$.

{\it Time-limited:} If we wish to restrict the growth in the 
time requirement to the $q^{k/2}$ that applies to our original
technique, then we use the cut-off
$$
m+n \leq k = 3\bmax+2
$$
Again we use our original technique for  $m \leq \bmax$ and use
the pivoting algorithm for rectangles of size $ m = (a+b)$ by 
$n = (2b+1)$ for $a+b$ from $\bmax+1$ to $\lfloor k/2 \rfloor$

In the work presented here we have used the `time-limited' form.
The algorithm for evaluating finite lattices by `pivoting' is

\begin{itemize}
\item[{\sl 1:}] For each of the $q^{a}$ states of the fixed line
(the filled circles in \figpivot): 
\begin{itemize}
\item[{\sl 1.1:}] Construct the conditional lattice sum.
\item[{\sl 1.2:}] Multiply the conditional sum by the internal 
weight of the fixed line.
\item[{\sl 1.3:}] Add the product to the running total for the 
finite lattice partition function.

\item[]The procedure for building up the conditional sum is:
\begin{itemize}
\item[{\sl 1.1.1:}] For each $x$ from $1-a$ to 0, build up the 
lattice sum for the column  $(x,y)$ for $y$ from 1 to $b$. (Region
I of \figpivot.)
\item[{\sl 1.1.2:}] For each $x$ from $1$ to b, build up the lattice 
sum for the partial column  $(x,y)$ for $y$ from b to $x$. (Region
II of \figpivot.)
\item[{\sl 1.1.3:}] For each $y$ from $b-1$ to 1, build up the lattice 
sum for the partial row  $(x,y)$ for $x$ from $b$ to $y$. (Region III
of \figpivot.)
\item[{\sl 1.1.4:}] Build up the lattice sum
for the row  $(x,0)$ for $x$ from $b$ to $1$ (IV).
\item[{\sl 1.1.5:}] For each $y$ from $-1$ to  $1-b=-c$, build up 
the lattice sum for the partial row  $(x,y)$ for $x$ from $b$ to $-y$. 
(Region V in \figpivot.)
\item[{\sl 1.1.6:}] If $b = c$, build up the lattice sum
for the diagonal  $(x,-x)$ for $x$ from $b$ to $1$. (Not present in
\figpivot.)
\item[{\sl 1.1.7:}] For each $x$ from $b-1$ to 1, build up the lattice 
sum for the partial row  $(x,y)$ for $y$ from $-b$ to $-x$. 
(Region VII of \figpivot.)
\item[{\sl 1.1.8:}] For each $x$ from $0$ to $1-b$, build up the 
lattice sum for the partial column $(x,y)$ for $y$ from $-b$ to $-1$.
(Region VIII of \figpivot.)
\end{itemize}
\end{itemize}
\end{itemize}

Another new feature of our calculations is the choice of a machine
(Cray YMP--EL) which emphasises processing speed rather than large
memory. The fact that most of our basic operations in the transfer
matrix approach are actually operations on truncated series
means that we can readily utilise the vector capabilities of such a
machine by ensuring that the series operations are performed
in vector mode. We note that the pivoting algorithm also permits
a high degree of parallel operation since each of the sums for
a given centre line can be evaluated independently of the others.

In order to deal with the large integer coefficients in the
series the calculations were performed using modular arithmetic
(see, e.g., Knuth 1969). Utilising the standard 46-bit integers 
of the Cray we used the set of primes, $p_{i} = 2^{23}-x_{i}$ with 
$x_{i} \in \{15,21,27,37,61,\ldots \}$. We had to use four primes 
for the spin-1 calculations and five primes for the 
spin-$\frac{1}{2}$ calculations. Each run with $\bmax=8$ 
($\bmax=12)$ for the spin-1 (spin-$\frac{1}{2}$) model required 
approximately 63 (48) CPU hours. 

\section{Expansions}
\setcounter{equation}{1}

For the spin-1 Ising model in a homogeneous magnetic field $h$ we
write the Hamiltonian as

$$
{\cal H} = \sum_{\langle ij \rangle} J(1-S_{i}S_{j}) +
           \sum_{i}h(1-S_{i})
\eqno(\thesection.\theequation) 
\addtocounter{equation}{1}
$$

where the spin variables $S_{i} = 0,\pm 1$.
The first sum is over all nearest neighbour pairs on the
square lattice and the second sum is over all sites. The constants
are chosen so that the $S_{i} =1$ ground state has zero energy. 
The low-temperature expansion, as described by Sykes and Gaunt
(1973), is based on perturbations from the $S_{i} = 1$ ground state.
The expansions are obtained in terms of the low-temperature variable
$u = \exp(-\beta J)$ and the field variable $\mu = \exp(-\beta h)$,
where $\beta = 1/kT$. The expansion of the partition function in 
powers of $u$ may be expressed as

$$
Z = \sum_{k=0}^{\infty} u^{k}\Psi_{k}(\mu) = 
1 + u^{4}\mu + u^{7}\mu^{2} + \ldots
\eqno(\thesection.\theequation) 
\addtocounter{equation}{1}
$$

where $\Psi_{k}(\mu)$ are polynomials in $\mu$. We express the
field variable as $\mu = 1-x$ and truncate the field dependence at
$x^{2}$ and thus find

$$
Z = Z_{0}(u) + xZ_{1}(u) + x^{2}Z_{2}(u) + \ldots.
\eqno(\thesection.\theequation) 
\addtocounter{equation}{1}
$$

According to standard definitions the order parameter,
or the spontaneous magnetisation, is the derivative of the 
free energy, $F= -kT\ln Z$, with respect to $h$,

$$
M(u) = M(0) + \frac{1}{\beta}\left. 
       \frac{\partial \ln Z}{\partial h}\right|_{h=0}
     = 1+ Z_{1}(u)/Z_{0}(u),
\eqno(\thesection.\theequation) 
\addtocounter{equation}{1}
$$

since $x=0$ in zero field. For the susceptibility we find

$$
\chi(u) = \left. \frac{\partial M(h)}{\partial h}\right|_{h=0} =
\frac{\partial}{\partial h}\left. \left(
Z^{-1}\frac{\partial Z}{\partial h}\right)\right|_{h=0} =
\beta \left[2\frac{Z_{2}(u)}{Z_{0}(u)}-\frac{Z_{1}(u)}{Z_{0}(u)}-
\left( \frac{Z_{1}(u)}{Z_{0}(u)} \right)^{2}\right].
\eqno(\thesection.\theequation) 
\addtocounter{equation}{1}
$$

The specific heat series is derived from the zero field partition
function (via the internal energy 
$U= -(\partial/\partial \beta) \ln Z_{0}$),

$$
C_{v}(u) = \frac{\partial U}{\partial T} = 
\beta^{2}\frac{\partial^{2}}{\partial \beta^{2}}\ln Z_{0}=
(\beta J)^{2}\left( u\frac{\mbox{d}}{\mbox{d}u} \right)^{2} 
\ln Z_{0}(u).
\eqno(\thesection.\theequation) 
\addtocounter{equation}{1}
$$

The resulting series for $M(u)$, $\beta^{-1}\chi(u)$, and
$(\beta J)^{-2}C_{v}(u)$ are given in \tablespin1ser. The number 
of terms derived correctly with the finite lattice
method is given by the power of the lowest-order connected graph
not contained in any of the rectangles considered. Since we are
using the {\em time-limited} cut-off the simplest such graphs
are chains of $3\bmax+2 = r$ sites all in the '0' state. From
the spin-1 Hamiltonian we see that such chains give rise to terms
$u^{3r+1}$. The series are thus correct to order $u^{3r}=u^{9\bmax+6}$.
We have checked this explicitly by calculating the series for
$\bmax = 2, 3\ldots, 7$ and checking that the terms through 
$u^{9\bmax+6}$ agree with the final 79-term series derived using 
$\bmax=8$. An additional spin-1 coefficient was calculated by a 
correction procedure explained in the Appendix. These new 
series are significant extensions to the hitherto longest series 
(45 terms) due to Adler and Enting (1984).

By the same methods {\em mutatis mutandis}, we calculated, with
$\bmax = 12$, a new 78-term series for the spin-$\frac{1}{2}$ 
Ising model in the low-temperature expansion variable 
$u=\exp(-2\beta J)$. Note that the lowest order graphs
not counted are chains of spins flipped with respect to the
ground state. But now only {\em broken} bonds pick up a factor
of $u$. With chains of length $r$ there are $2r+2$ broken bonds
so the series are correct to order $u^{2r+1}=u^{6\bmax+5}$.
The resulting series are listed in \tablespinhalfser. The
spin-$\frac{1}{2}$ magnetisation and specific heat series are of
course known exactly, so only the susceptibility series is new. We
nevertheless list the coefficients of all three quantities, partly for
completeness and partly for verification of our algorithm.

\section{Analysis of the spin-1 series}
\setcounter{equation}{1}

The series for the spontaneous magnetisation, the susceptibility and
the specific heat of the spin-1 Ising model are expected to exhibit
critical behaviour of the forms

$$
M(u) \sim A_{M}(u_{c}-u)^{\beta}[1+a_{M,1}(u_{c}-u)^{\Delta_{1}}+
          b_{M,1}(u_{c}-u) + \ldots ],
\eqno(\thesection.\theequation) 
\addtocounter{equation}{1}
$$
$$
\chi(u) \sim A_{\chi}(u_{c}-u)^{-\gamma'}[1+
          a_{\chi,1}(u_{c}-u)^{\Delta_{1}}+
          b_{\chi,1}(u_{c}-u) + \ldots ],
\eqno(\thesection.\theequation) 
\addtocounter{equation}{1}
$$
$$
C_{v}(u) \sim A_{C}(u_{c}-u)^{-\alpha'}[1+
          a_{C,1}(u_{c}-u)^{\Delta_{1}}+
          b_{C,1}(u_{c}-u) + \ldots ].
\eqno(\thesection.\theequation) 
\addtocounter{equation}{1}
$$

By universality it is expected that the leading critical exponents 
equal those of the spin-$\frac{1}{2}$ Ising model, i.e.,
$\beta=\frac{1}{8}$, $\gamma' = \frac{7}{4}$ and $\alpha' = 0$
(logarithmic divergence). One of the major differences between the 
two models appears to be that the non-analytic confluent terms 
aren't present in the spin-$\frac{1}{2}$ model (the $a_{1}$'s 
equal zero). 

\subsection{$u_{c}$ and the leading critical exponents}

The low-temperature spin-1 series is ill-behaved because there are
non-physical singularities closer to the origin than the physical
singularity, thus rendering ratio methods useless. The series may
still be analyzed using differential 
approximants (Guttmann 1989), which provides an effective analytic
continuation beyond the radius of convergence, thus allowing accurate
estimation of critical parameters even when the dominant singularity
is non-physical. It is also often useful to change the series 
variable by a transformation leading to a new series in which the
singularity closest to the origin is the physical one. However,
such 'singularity-moving' transformations may introduce
long-period oscillations (Guttmann 1989) seriously impairing the
accuracy of ratio methods. 

It turns out that ordinary Dlog Pad\'{e} approximants, 
equivalent to first order homogeneous differential approximants, 
works best for the magnetisation series. By averaging over several 
[N/M] approximants with $|N-M| \leq 4$ using at least 65 terms
of the series ($N+M > 64$) we find the following estimates for the
critical point $u_{c} = 0.554075(15)$ and exponent $\beta = 0.1253(3)$. 
The number in parenthesis is the error in the last digit(s) given as 
three standard deviations. We find that approximants using fewer than
$\simeq 60$ terms deviate systematically from these averages.
In \figbetavnt\ we have plotted the estimates for $\beta$ {\em vs}
the number of terms ($N+M+1$) from the series utilized in the Dlog
Pad\'{e} approximants. We clearly see how the $\beta$-estimates 
settle down to a plateau around $\beta \simeq 0.1253$ when more
than 60 terms are used. The estimate for $\beta$ is slightly higher
than the expected exact value $\beta = \frac{1}{8}$. In addition to
the physical singularity at $u_{c}$, we find that the magnetisation
series has a singularity on the negative $u$-axis at 
$u_{-} = -0.59853(4)$ with exponent $\beta_{1} = 0.1247(6)$ and
a pair of complex roots at $u_{\pm} = -0.30183(5)\pm 0.37870(4)i$
with exponent $\beta_{\pm} = -0.127(3)$. Note that the non-physical
singularity $u_{\pm}$ is closer to the origin than the physical
singularity $u_{c}$. First order {\em inhomogeneous} and second 
order differential approximants do not work very well for the 
magnetisation series, as evidenced by error estimates which are 
generally at least an order of magnitude larger than 
in the simple Dlog Pad\'{e} case. If we assume that the exact value
of $\beta = \frac{1}{8}$ we have to change our estimate of $u_{c}$.
We find basically a linear relationship between the estimates for
$\beta$ and $u_{c}$, and for $\beta=\frac{1}{8}$ we find 
$u_{c} = 0.554065(5)$. We have obtained a very similar result by
analyzing the series for $M^{8}(u)$ using ordinary ({\em not} Dlog)
Pad\'{e} approximants. Raising the magnetisation series to the 
8th power and looking for simple zeros and poles of the resulting
series obviously corresponds to biasing the magnetisation series
to have a leading critical exponent of $\frac{1}{8}$. We find
that the function given by this series has zeroes at 
$u_{c} = 0.554063(10)$ and $u_{-} = -0.598555(10)$ plus a conjugate 
pair of simple poles at $u_{\pm} = -0.30198(3) \pm 0.37857(5)i$. 
For comparison we note that the spontaneous magnetisation of the 
quadratic spin-$\frac{1}{2}$ Ising model is given by the formula
(Onsager 1944, Yang 1952)

$$
I(u)=\left[\frac{1+u^{2}}{(1-u^{2})^{2}}
                (1-6u^{2}+u^{4})^{1/2}\right]^{1/4},
$$

from which we see that the magnetisation have singularities
with exponent $\frac{1}{8}$ at $\pm(\sqrt{2}-1)$ and $\pm(\sqrt{2}+1)$,
with exponent $\frac{1}{4}$ at $\pm i$ and finally with exponent 
$-\frac{1}{2}$ at $\pm 1$.  

The success of ordinary Dlog Pad\'{e} approximants in analyzing
the magnetisation series stems from the absence of analytic 
background terms. In the
susceptibility and specific heat series such background terms
are indeed present and obscure the leading critical behaviour.
However, inhomogeneous differential approximants are generally
successful in dealing with such terms. In \tablegammauc\ we have 
listed the estimates for $\gamma'$ and $u_{c}$ obtained by averaging 
many different approximants to the susceptibility series. We find
that ordinary Dlog Pad\'{e} approximants (the first order approximants 
with $L=0$) yield quite stable estimates but that the estimate
for $u_{c}$ is quite a bit larger than for the magnetisation, and
that $\gamma'$ is markedly larger than the expected exact value
$\gamma' = \frac{7}{4}$. However, once the order of the inhomogeneous
polynomial is larger than 2, the estimates for $\gamma'$ becomes
fully consistent with the expected behaviour, indeed we see that
the first order approximants favor a value a little larger than 
$\frac{7}{4}$ whereas the second order approximants favor a value
slightly below $\frac{7}{4}$. Taken together there seems little
doubt the exact value indeed is $\gamma' = \frac{7}{4}$. Again 
assuming a linear relationship between $\gamma'$ and $u_{c}$, we
find that $u_{c} \simeq 0.554065$. The estimates for
the critical exponent $\gamma'$ exhibit the same trend as those
for $\beta$, i.e., when fewer than $\simeq 60$ terms are used the
estimates are generally clearly $> \frac{7}{4}$ with larger 
deviations when fewer terms is involved in estimating $\gamma'$.
When more than 60 terms are used, the estimates reach a plateau
around a value $\simeq 1.755$, but with a spread that clearly
includes the expected exact value $\gamma' = \frac{7}{4}$.
In this case we find additional singularities at $u_{-} = -0.5984(1)$ 
with exponent $-1.725(15)$ and $u_{\pm} = -0.30194(2)\pm 0.37877(2)i$ 
with an exponent of $-1.175(10)$. A closer examination of the various 
approximants revealed that as the estimates of $u_{-}$ approach 
the value found from the magnetisation series, $u_{-} \sim -0.598555$, 
the exponent approaches $-1.75$. It is thus very likely that the 
exponents at $u_{c}$ and $u_{-}$ are equal. 

The analytic background term is stronger in the specific heat
series as can be seen in \tablealphauc\ where we have listed the
estimates for $\alpha'$ and $u_{c}$. The first order approximants
yield no useful results with $L=0,1$. Once the order of the 
inhomogeneous polynomial becomes larger than 3 the first order 
approximants clearly yield an estimate consistent with $\alpha' = 0$. 
This time a linear relationship between $\alpha'$ and $u_{c}$ indicates
$u_{c} \simeq 0.554070$ when $\alpha' =0$. In addition we find a pair 
of complex roots at $u_{\pm} = -0.301945(15) \pm 0.378776(10)$ with an 
exponent (divergence) of $-1.172(10)$. These conclusions are fully 
confirmed by the results of the analysis using second order 
differential approximants. There is also evidence for a singularity at 
$u_{-} = -0.598(6)$, but as can be seen from the size of the error
esimate it is not well-defined. This is also reflected in the 
estimates of the associated exponent, ranging from 0.5 to -0.5, 
with values close to zero when $u_{-} \sim -0.5985$. This could 
indicate a logarithmic singularity at $u_{-}$, though the evidence 
is very weak. A stronger case can be made by looking at the series 
for the derivative of the specific heat, d$C_{v}(u)$/d$u$, which 
should have simple poles at $u_{c}$ and $u_{-}$ if $C_{v}(u)$ has
logarithmic singularities at these points. A Dlog Pad\'{e} analysis
of the series revealed singularities at $u_{c} = 0.5540(5)$ with
exponent $-1.00(4)$, at $u_{-} = -0.5975(10)$ with exponent
$-0.95(5)$, and at $u_{\pm} = -0.30195(1)\pm0.37878(1)i$ with
exponent $-2.177(15)$. These results thus confirm the results
from the analysis of the specific heat series itself.

The scaling law, $\alpha' +2\beta +\gamma' = 2$, is seen to hold at 
both the critical point $u_{c}$ and at the non-physical singularity
$u_{-}$. Likewise, for the spin-$\frac{1}{2}$
Ising model this scaling law holds at the sigularities
$\pm(\sqrt{2}-1)$ since $\alpha' =0$, $\beta=\frac{1}{8}$, and
$\gamma'=\frac{7}{4}$ in both cases.  At the other non-physical 
singularity $u_{\pm}$ we find $\alpha' + 2\beta + \gamma' = 2.09(3)$ 
for the spin-1 model. From the exact solutions for the zero-field 
partition function and spontaneous magnetisation of the 
spin-$\frac{1}{2}$ Ising model it follows that $\alpha' =0$ and 
$\beta=\frac{1}{4}$ at $u_{\pm} = \pm i$. A differential approximant
analysis of the susceptibility series yields the estimate 
$\gamma' = 1.555(10)$ at $u_{\pm}$. So for the spin-$\frac{1}{2}$
we find, at $u_{\pm}$, that $\alpha'+2\beta+\gamma' = 2.055(10)$.
It seems highly likely that $\alpha' +2\beta +\gamma' = 2$ holds
at all the non-physical singularities. This would mean that the 
exponent corresponding to $\gamma'$ at $u=u_{\pm}=\pm i$ for the
spin-$\frac{1}{2}$ Ising model would be $\frac{3}{2}$ exactly.
For the spin-1 Ising model the situation is less clear. At $u=u_{\pm}$,
the analogue of $\alpha'$ is 1.172, and the analogue of $\gamma'$ is
1.175. It is possible that they are both $\frac{9}{8}$ exactly, or
that one is 1 and the other is $\frac{5}{4}$. We have not been able 
to make these results more precise.

As mentioned above ratio methods are of use only if one can
find a transformation that maps the non-physical singularity
outside the transformed physical disc. One such transformation
is given by $u=x/(2-x)$, which leads to a new (high-temperature like)
expansion variable, $x= 1-\tanh (\beta J/2)$. Among the various 
extrapolation methods (Guttmann 1989) we find that the best 
over-all convergence is obtained from the Neville-Aitkin table.
From the magnetisation and susceptibility series we obtain 
the estimates $1/x_{c} = 1.4024(3)$, $\beta=0.12(1)$ and 
$\gamma' = 1.75(1)$. The exponent estimates are from biased approximants
using the accurate value $x_{c} = 0.71305(1)$ obtained from the
differential approximant analysis. While this type of analysis yields 
estimates of lesser accuracy than the analysis based on differential 
approximants it is nevertheless reassuring that the two methods are in 
agreement. Other extrapolation methods generally yield similar
though less accurate estimates. The major source of error in all 
the methods is the presence of long-period oscillations in 
the extrapolations. 

\subsection{The critical amplitudes}

We have calculated the critical amplitudes using two different
methods, both of which are very simple and easy to implement.
In the first method, we note that if $f \sim A(1-u/u_{c})^{-\lambda}$,
then it follows that 
$(u_{c} -u)f^{1/\lambda}|_{u=u_{c}} \sim A^{1/\lambda}u_{c}$. So we
simply form the series for $g(u) = (u_{c} -u)f^{1/\lambda}$ and
evaluate Pad\'{e} approximants to this series at $u_{c}$. The result
is just $A^{1/\lambda}u_{c}$. This procedure works well for the
magnetisation and susceptibility series (it obviously cannot
be used to analyse the specific heat series). For the magnetisation
we find that the spread of various approximants is minimal at
$u_{c} = 0.554063$ where $A_{M} = 1.208496(4)$. Allowing for a
value of $u_{c}$ between 0.55406 and 0.55407 we find
$A_{M} = 1.2084(2)$. A similar analysis for the susceptibility
yields the closest agreement at $u_{c} = 0.554065$ with
$A_{\chi} = 0.06164(1)$. Again allowing for a wider choice in $u_{c}$
we find $A_{\chi} = 0.0616(2)$.

In the second method, proposed by Liu and Fisher (1989), one
starts from $f(u) \sim A(u)(1-u/u_{c})^{-\lambda}+B(u)$ and then
forms the auxiliary function  $g(u) = (1-u/u_{c})^{\lambda}f(u) 
\sim A(u) + B(u)(1-u/u_{c})^{\lambda}$. Thus the required amplitude
is now the {\em background} term in $g(u)$, which can be obtained
from inhomogeneous differential approximants (Guttmann (1989) p89).
In \tableampl\ we have listed the estimates obtained by averaging
over various first order differential approximants using at least
65 terms of the series with $u_{c} = 0.554065$. 
The results for the magnetisation $A_{M} = 1.2090(20)$ and the
susceptibility $A_{\chi} = 0.0625(10)$ agree with those obtained
above, though the error estimates are much larger. These results
are not seriously affected by allowing for a wider choice of $u_{c}$.
This method can also be used to study the specific heat series. One
now starts from $f(u) \sim A(u)\ln (1-u/u_{c})+B(u)$ and then looks
at the auxiliary function $g(u) = f(u)/\ln (1-u/u_{c})$. As before
the amplitude can be obtained as the background term in $g(u)$.
The results of the analysis are listed in \tableampl\ from which we
get the final estimate $A_{C} = 19.75(50)$.

Jugding from the error estimates it would seem that the first method
for calculating the amplitudes is superior to the second method. 
This apparent superiority does however not hold up under further
scrutiny. We checked the two methods on the spin-$\frac{1}{2}$ 
susceptibility series where the leading amplitude has been calculated
to high accuracy (Wu \etal 1976). In the widely accepted standard 
notation (Fisher 1967), $T\chi = C_{0} |1-T/T_{c}|^{-7/4}$, one has 
to 10 decimal places $C_{0} = 0.0255369719...$. In our analysis we
assumed a singularity, $\chi(u) \sim A_{\chi}|1-u/u_{c}|^{-7/4}$. 
With $u=\exp(-2\beta J)$ it follows that 
$C_{0}=4u_{c}^{4}(-\ln u_{c})^{-7/4}A_{\chi}$, where the factor
$u_{c}^{4}$ arise because we analyse the series for $\chi(u)/u^{4}$,
the factor $(-\ln u_{c})^{-7/4}$ is caused by the change of variable
and the factor 4 is a matter of definition. Since,
$u_{c} = \sqrt{2} -1$, we find that $A_{\chi} = 0.584850251...$. 
Using the two methods to calculate $A_{\chi}$ we get the estimates
$A_{\chi} = 0.58488(1)$ from the first method and 
$A_{\chi} = 0.58490(5)$ from the second method. This clearly shows
that the smaller error estimate from the first method cannot be
taken too seriously as both estimates are only marginally 
consistent with the exact result.

We thus conclude that $A_{M} = 1.208(4)$, $A_{\chi}=0.0615(2)$ and
$A_{C} = 19.8(1.0)$. Note that these amplitudes are obtained by
analyzing the series for $M(u)$, $\beta^{-1}u^{-4}\chi(u)$, and
$(\beta J)^{-2} u^{-4}C_{v}(u)$, respectively, assuming in each
case a singularity $\propto |1-u/u_{c}|^{\lambda}$. Changing to
the standard notation (Fisher 1967) and getting rid of the various
prefactors we find: The amplitude of $M(T)$ is 
$B= (-\ln u_{c})^{1/8}A_{M}= 1.131(4)$, the amplitude of
$T \chi(T)$ is $C_{-} = (-\ln u_{c})^{-7/4}u_{c}^{4}A_{\chi} =
0.0146(5)$ and finally the amplitude of $C_{v}(T)$ is
$A_{-} = (-\ln u_{c})^{2}u_{c}^{4}A_{C} = 0.65(3)$. From this we
find the Watson invariant (Watson 1969)

$$
   A_{-}B^{-2}C_{-} = 0.0074(6),
$$

which should be independent of the choice of lattice.

\subsection{The confluent exponent}

We have studied the series using three different methods in order 
to estimate the value of the confluent exponent. In the first method,
due to Baker and Hunter (1973), one transforms the function $F$,

$$
F(u) = \sum_{i=1}^{N}A_{i}
\left( 1-\frac{u}{u_{c}}\right)^{-\lambda_{i}} 
= \sum_{n=0}^{\infty}a_{n}u^{n}
\eqno(\thesection.\theequation) 
\addtocounter{equation}{1}
$$
into an auxiliary function with simple poles at $1/\lambda_{i}$.
We first make the substitution $u = u_{c}(1-e^{-\zeta})$ and find
$$
F(u(\zeta)) = \sum_{i=1}^{N}A_{i}\exp\left[ -\lambda_{i}\ln\left(
1-\frac{u}{u_{c}}\right) \right] = 
\sum_{i=1}^{N}A_{i}e^{\lambda_{i}\zeta} =
\sum_{i=1}^{N}\sum_{k=0}^{\infty} 
\frac{A_{i}\lambda_{i}^{k}\zeta^{k}}{k!}
\eqno(\thesection.\theequation) 
\addtocounter{equation}{1}
$$
By multiplying the coefficient of $\zeta^{k}$ by $k!$ we get the
required auxiliary function
$$
{\cal F}(\zeta) = \sum_{i=1}^{N}\sum_{k=0}^{\infty}
A_{i}(\lambda_{i}\zeta)^{k} = 
\sum_{i=1}^{N}\frac{A_{i}}{1-\lambda_{i}\zeta},
\eqno(\thesection.\theequation) 
\addtocounter{equation}{1}
$$
which has poles at $\zeta = 1/\lambda_{i}$, with residues at the 
poles of $-A_{i}/\lambda_{i}$. In \tablebetadelta\ we have listed
the estimates for the leading critical exponent $\beta$ and the 
confluent exponent $\Delta_{1}$ and their corresponding amplitudes
as obtained from an analysis of the Baker-Hunter transformed 
spontaneous magnetisation series
with $u_{c} = 0.554065$. Only the [N/N+1] approximants yield 
useful information. The [N+1/N] approximants does not even find
a pole close to $-8 = -1/\beta$ whereas some of the [N/N]
approximants find a pole at $-8.4(2)$ but no corresponding
estimate for the confluent exponent. The results of the
[N/N+1] approximants certainly confirm
that the leading critical exponent is $\frac{1}{8}$, and the
corresponding estimates of the critical amplitudes $A_{M}$ are also 
in excellent agreement with the results obtained in the previous
section. The estimates for $\Delta_{1}$ are less stable, but the
approximants with $N \geq 34$ are consistent with an estimate
$\Delta_{1} = 1.06(2)$. These conclusions are unaltered by
looking at $u_{c} = 0.554060$ and 0.554070 though we note that
by far the best agreement between different approximants is 
for $u_{c} = 0.554065$. The results for the susceptibility series
are listed in \tablegammadelta\ for $u_{c} =0.554065$. In this case
we again confirm the values of the leading critical exponent and
critical amplitudes. The results for the confluent singularity
is much more confusing as the [N/N+1] approximants yields the
estimate $\Delta_{1} = 1.155(5)$ very different from the estimate
$\Delta_{1} = 1.4(1)$ obtained from the [N+1/N] approximants. The
[N/N] approximants generally yield no estimates for this value
of $u_{c}$. However, at slightly higher values of $u_{c}$,
the [N/N] approximants also become useful, though they favor neither
one nor the other of the $\Delta_{1}$-estimates cited above.
This is clearly seen in \figsusbht\ where we have plotted the
estimate for $\Delta_{1}$ {\em vs} the parameter $u_{c}$ used in
the Baker-Hunter transformation for various approximants with
$N \geq 36$. The different approximants [N/N+1], [N/N], and [N+1/N] 
clearly bunch together in three distinct classes. The [N/N+1]
approximants exhibit a narrow crossing at $\Delta_{1} = 1.13(2)$
and $u_{c} = 0.55410(5)$ whereas the [N+1/N] approximants cross
at $\Delta_{1} = 1.40(5)$ and $u_{c} = 0.55406(2)$ and the
[N/N] approximant though not intersecting mutually do seem to
cross through both the above regions. Note that the [N/N] approximants
doesn't yield any estimates below $u_{c} \simeq 0.55406(2)$.
From these results it is not possible to infer a final estimate
for $\Delta_{1}$ and we are unable to explain the very different
behaviour exhibited by the various sets of approximants.

The second method, due to Adler \etal (1981), involves studying
Dlog Pad\'{e} approximants to the function $G(u)$, where
$$
G(u) = \lambda F(u) + (u_{c}-u)\mbox{d}F(u)/\mbox{d}u.
$$
The logarithmic derivative to $G(u)$ has a pole at $u_{c}$ with
residue $\lambda + \Delta_{1}$, where $\lambda$ is the leading 
critical exponent. We evaluate the Dlog Pad\'{e} approximants
for a range of guesses for $u_{c}$ and $\lambda$. For each such 
guess we thus find an estimate for $\Delta_{1}$; for the correct
value of $u_{c}$ and $\lambda$ we should see a convergence region
in the $(u_{c},\lambda,\Delta_{1})$-space. In practice we allways
froze $\lambda$ at its expected exact value and plotted
$\Delta_{1}$ as a function of $u_{c}$. \figmagucamp\ 
shows $\Delta_{1}$ as a function of $u_{c}$ with $\beta = \frac{1}{8}$
using the spontaneous magnetisation series. In this 
figure we see a very narrow convergence region at 
$u_{c} = 0.544066(1)$ and $\Delta_{1} = 1.06(2)$. In \figsusucamp\ 
we have plotted similar results from an analysis of the susceptibility
series with $\gamma' = \frac{7}{4}$. In this case we find a narrow 
crossing region at $u_{c} = 0.5540695(15)$ and $\Delta_{1}=1.15(2)$.

The last method, also due to Adler \etal (1983), is a generalisation
of an approach devised by Roskies (1981) to study the high-temperature
susceptibility of the three-dimensional spin-$\frac{1}{2}$ Ising model.
The first step is to transform the series $F(u)$ to ones in

$$
y = 1-(1-u/u_{c})^{\Delta},
$$

where $u_{c}$ is assumed known but $\Delta$ is a variable parameter.
Next one looks at different Pad\'{e} approximants to the function

$$
{\cal G}_{\Delta} = \Delta (y-1)
\frac{\mbox{d}}{\mbox{d}y}\ln F(y,\Delta) 
\sim \lambda +O[(1-y)^{\Delta_{1}/\Delta}].
$$

For the correct guesses for $u_{c}$ and $\Delta_{1}$ the various
Pad\'{e} approximants should intersect and give a correct estimate
for the leading critical exponent $\lambda$. In \figbetagrt\ we have
plotted the estimate for $\beta$ as a function of the transformation
variable $\Delta$ with $u_{c} = 0.554065$ using the spontaneous
magnetisation series. A narrow crossing region is found at
$\Delta_{1} = 1.11(1)$ and $\beta = 0.12490(5)$. \figgamma'grt\ shows
a similar plot but for the susceptibility series; in this case we 
locate the crossing at $\Delta_{1} = 1.16(1)$ and $\gamma' = 1.749(1)$.

The value of the correction-to-scaling exponent of two-dimensional Ising
systems has been addressed by several authors. Nienhuis (1982) has 
mapped the $q$-state Potts model in two dimensions onto a model which is
in the Ising universality class (when $q$ is set to 2). In that case one
obtains a value $\Delta_{1} = 4/3$. Barma and Fisher (1985) obtained
$\Delta_{1} = 1.35 \pm 0.25$ at the pure Ising critical point of the
Klauder and Double Gaussian models. They point out that while this is
inconsistent with the expectation of pure logarithmic corrections to
scaling, it is possible for these logarithmic terms to have amplitudes 
that vanish on approach to the pure Ising limit in such a way as to be
describable by an apparent correction-to-scaling exponent $\Delta_{1}=4/3$.

The various estimates we have obtained for the spin-1 model appear to
be entirely consistent with this complex and subtle behaviour.

\section{Summary and Discussion}

In this work we have substantially extended the existing 
low-temperature series for the two-dimensional spin-1 Ising model 
by introducing a new version of the finite lattice method which 
saves considerable space over previous implementations. The 
usefulness of such extended series is demonstrated by our analysis, 
in which we find that the correct results are clearly evident only 
after more than 60 terms in the series!  

The analysis of the new spin-1 series provides us with an accurate
estimate of the critical point, $u_{c} = 0.554065(5)$, where the
error estimate is chosen rather conservatively, and reflects the
descripancy between estimates obtained from the three series
and using different methods of analysis. Our value is in
excellent agreement with a recent estimate $u_{c} \simeq 0.554066$
obtained by Lipowski and Suzuki (1992) from a transfer-matrix version
of the double cluster mean-field approximation, and also agrees
with the estimate $u_{c} = 0.554071(3)$ of Bl\"{o}te and Nightingale
(1985) obtained by phenomenological renormalization using the
correlation length of a $n\times\infty$ systems calculated with
transfer matrix techniques. However, our results and those 
mentioned above, clearly rule out a recent conjecture
$u_{c} = 0.553887\ldots$ by Urumov (1988) for the {\em exact} value
of the critical point obtained from a generalisation of a method
proposed by \v{S}vraki\'{c} (1980) to calculate the exactly known
equations for the critical temperature of the spin-$\frac{1}{2}$
Ising model on a chequerboard lattice. 

The evidence for the leading critical exponents clearly show
that the spin-1 Ising model belongs to the usual Ising universality
class, a point also supported strongly by the results of Bl\"{o}te 
and Nightingale (1985), i.e., the critical exponents take the values 
$\beta=\frac{1}{8}$, $\gamma'=\frac{7}{4}$, and $\alpha' = 0$, exactly.
In addition we find a non-analytic confluent correction with an 
exponent $\Delta_{1} > 1$. Generally the magnetisation series
favors a value of $\sim 1.05$ whereas the susceptibility series
yields estimates $\sim 1.15$. One notable exception is the [N+1/N]
approximants to the Baker-Hunter transformed susceptibility series 
which leads to estimates $\sim 1.4$. From field-theoretic arguments
it is expected that the value of $\Delta_{1}$ should be the same
for both series. This leads us to the final estimate 
$\Delta_{1} = 1.1(1)$ in good agreement with the result 
$ 1 < \Delta_{1} < 1.3$ of Adler and Enting (1984), but clearly
smaller than the  field-theoretic prediction $\Delta_{1} = 1.4$ 
(Le Guillou and Zinn-Justin, 1980), and the "exact" value
$\Delta_{1} = 4/3$ found by Nienhuis (1982). 

The new finite lattic method is also directly generalisable to 
higher spin Ising models, as well as to other two-dimensional 
systems. The extended series for the low-temperature susceptibility 
of the quadratic spin-$\frac{1}{2}$ Ising model is in excellent 
agreement with the numerical predictions of Gartenhaus and McCullough 
(1988). Our exact coefficients agree to all 7 predicted significant 
digits.

{\Large \bf Acknowledgements}

We would like to thank Michael E. Fisher for helpful discussions on
the question of correction-to-scaling exponents.
Financial support from the
Australian Research Council is gratefully acknowledged by AJG and IJ.

\newpage
{\Large \bf Appendix: First correction term}

In Section 2 we described how to obtain series expansions
from the finite lattice method. We ran the program upto
$b_{max}= 8$ yielding a series correct through order $u^{78}$.
When running the program one may actually calculate the truncated
series in $u$ to orders beyond that to which the series coefficients
are correct. One would expect at least the first incorrect
coefficient derived in this fashion to deviate only a little from 
the correct one. In this Appendix we describe a correction procedure 
whereby we were able to obtain one additional correct term and thus
extend the series to order $u^{79}$. At first we just looked at
the first correction term for $\bmax = 2-7$, i.e., the 
difference between the correct series coefficients and the first
incorrect ones for a given value of $\bmax$, but were unable
to find a pattern allowing us to 'guess' the correction term for
$\bmax=8$. Next we reran the program deriving a series using
$q=2$ hoping that the correction terms for the $q=2$ case were
the same as those for the spin-1 model ($q=3$). It turns out that
the correction terms is the two cases are almost identical. 
In \tabcorr\ we have listed the amount by which the $q=2$ correction
terms fail to predict the spin-1 correction terms. One could
thus obtain almost correct spin-1 correction terms for $\bmax=8$
just using the $q=2$ correction term. However, we found that
the numbers in \tabcorr\ obey recursive relations. Let 
$\Delta_{x}^{b} = \delta_{x}^{b}/\bmax$, then we find the
following recurrence relations for the $\Delta$'s, in the
case of the partition function

$$
\Delta_{Z_{0}}^{b+2} -2\Delta_{Z_{0}}^{b+1}+\Delta_{Z_{0}}^{b} = -9,
\eqno(A2.1)
$$
the magnetisation
$$
\Delta_{M}^{b+3} -3\Delta_{M}^{b+2}+3\Delta_{M}^{b+1}
-\Delta_{M}^{b} = 81,
\eqno(A2.2)
$$
and the susceptibility
$$
\Delta_{\chi}^{b+4} -4\Delta_{\chi}^{b+3}+6\Delta_{\chi}^{b+2}
-4\Delta_{\chi}^{b+1}+\Delta_{\chi}^{b} = -972.
\eqno(A2.3)
$$

Using these recurrence relations we find $\Delta_{x}^{8}$ and
thus using the $q=2$ correction terms for $\bmax=8$ we can
finally find the correction terms for the spin-1 model and
thus the correct coefficient to $u^{79}$.

Though the correction procedure is rather cumbersome it should
be noted that the CPU-time required to calculate the truncated 
series with $\bmax=2-7$ for $q=3$ and $\bmax=2-9$ for $q=2$ is 
insignificant compared to the time it takes to run the
program with $\bmax=8$ and $q=3$.

\newpage

\hoffset=-0.5cm
\setlength{\parindent}{-0.5cm}

{\Large \bf References}

\refjl{Adler J and Enting I G 1984}{\JPA}{17}{2233}

\refjl{Adler J, Moshe M and Privman V 1981}{\JPA}{14}{L363}

\refbk{\dash 1983 in}{Percolation Structures and Processes}
{ed G Deutsher, R Zallen and J Adler {\em Ann. Israel Phys. Soc.}
vol 5 (Bristol: Adam Hilger)}

\refjl{Baker G A and Hunter D L 1973}{\PRB}{7}{3377}

\refjl{Barma M, Fisher M E 1985}{\PRB}{31}{5954}

\refjl{Bl\"{o}te H W J and Nightingale M P 1985}{Physica A}{134}{274}

\refjl{Enting I G 1978}{\JPA}{11}{563}

\refbk{\dash 1990 Algebraic techniques in latttice statistics:
The computational complexity of the finite lattice method}
{Third Australian Supercomputer Conference,}
{University of Melbourne, December, 1990: Proceedings}

\refjl{Fisher M E 1967}{Rep. Prog. Phys.}{30}{615}

\refjl{Fox P F and Guttmann A J 1973}{\JPC}{6}{913}

\refjl{Gartenhaus S and McCullough W S 1988}{\PRB}{38}{11688}

\refbk{Guttmann  A J 1989 Asymptotic Analysis of Power-Series
Expansions}{Phase Transitions and Critical Phenomena}
{vol 13, ed C. Domb and J. L. Lebowitz (Academic Press, New York).}

\refjl{Guttmann A J and Enting I G 1993}{\PRL}{70}{698}

\refbk{Knuth D E 1969}{Seminumerical Algorithms. The Art of
Computer Programming. Vol. 2}{(Addison-Wesley, Reading, Mass.).}  

\refjl{Le Guillou J C and Zinn-Justin J 1980}{\PRB}{21}{3976}

\refjl{Lipowski A and Suzuki M 1992}{J. Phys. Soc. Jpn.}{61}{4356}

\refjl{Liu A J and Fisher M E 1989}{Physica A}{156}{35}

\refbk{de Neef T 1975 Some Applications of Series Expansions in 
Magnetism}{Ph.D. Thesis}{ Technische Hogeschool Eindhoven }

\refjl{de Neef T and Enting I G 1977}{\JPA}{10}{801}

\refjl{Nienhuis B 1982}{\JPA}{15}{199}

\refjl{Onsager L 1944}{Phys. Rev.}{65}{117}

\refjl{Roskies R 1981}{\PRB}{24}{5305}

\refjl{\v{S}vraki\'{c} N M 1980}{Phys. Lett. A}{80}{43}

\refjl{Sykes M F, Essam J W, and Gaunt D S 1965}
{J. Math. Phys.}{6}{283}

\refjl{Sykes M F and Gaunt D S 1973}{\JPA}{6}{643}

\refjl{Urumov V 1988}{Phys. Stat. Sol. (b)}{145}{K59}

\refjl{Watson P G 1969a}{\JPC}{2}{1883, 2158}

\refjl{Wu T S, Mc Coy B M, Tracy C A and Barouch E 1976}{\PRB}{13}{316}

\refjl{Yang C N 1952}{Phys. Rev.}{85}{808}

\newpage

\hoffset=-1cm
\setlength{\parindent}{0cm}

{\Large \bf Figure Captions}

{\bf \figpivot:} The various regions of the lattice numbered
according to the order in which they are traversed
in the pivoting-algorithm. The $a$ spins on the filled circles
are fixed. In the case of this $12\times 16$ lattice 
we have $a = 4$, $b=8$ and $c=7$.

{\bf \figbetavnt:} Estimates for the leading critical exponent
$\beta$ of the spin-1 spontaneous magnetisation {\em vs} the
number of terms used by the Dlog Pad\'{e} approximants.

{\bf \figsusbht:} Estimates for the confluent exponent $\Delta$
{\em vs} the parameter $u_{c}$ of the Baker-Hunter transformation.
All Pad\'{e} approximants with $36 \leq N \leq 39$ are shown.

{\bf \figmagucamp:} The confluent exponent $\Delta$ {\em vs} $u_{c}$
as obtained from the method of Adler \etal (1981) applied to the
spontaneous magnetisation series.

{\bf \figsusucamp:} Same as in \figmagucamp\ but for the susceptibility
series.

{\bf \figbetagrt:} The exponent $\beta$ as a function of the parameter
$\Delta$ used in the generalised Roskie transformation of 
Adler \etal (1983) applied to the spontaneous magnetisation series.
The inset shows the details of the crossing region.

{\bf \figgamma'grt:} Same as in \figbetagrt\ but for the susceptibility
series.

\newpage

{\Large \bf Tables}

\vspace{1cm}

{\bf \tablespin1ser:} New low-temperature series for the spin-1 
\2D Ising magnetisation ($M(u) = \sum_{n}m_{n}u^{n}$), susceptibility
($\chi(u) = \sum_{n}x_{n}u^{n}$), and specific heat
($C_{v}(u) = \sum_{n}c_{n}u^{n}$).

\footnotesize
\begin{center}
\begin{tabular}{rrrr}
\hline \hline
 $n$ & \mbox{\hspace{25ex}$m_{n}$} & 
 \mbox{\hspace{27ex}$x_{n}$} & 
 \mbox{\hspace{28ex}$c_{n}$} \\
\hline
  0 &                             1
 &                             0
 &                             0 \\ 
  1 &                             0
 &                             0
 &                             0 \\ 
  2 &                             0
 &                             0
 &                             0 \\ 
  3 &                             0
 &                             0
 &                             0 \\ 
  4 &                            -1
 &                             1
 &                            16 \\ 
  5 &                             0
 &                             0
 &                             0 \\ 
  6 &                             0
 &                             0
 &                             0 \\ 
  7 &                            -4
 &                             8
 &                            98 \\ 
  8 &                             3
 &                            -6
 &                           -96 \\ 
  9 &                             0
 &                             0
 &                             0 \\ 
 10 &                           -30
 &                            90
 &                          1000 \\ 
 11 &                            48
 &                          -144
 &                         -1936 \\ 
 12 &                           -52
 &                           192
 &                          2064 \\ 
 13 &                          -120
 &                           480
 &                          5070 \\ 
 14 &                           368
 &                         -1372
 &                        -19012 \\ 
 15 &                          -612
 &                          2676
 &                         31950 \\ 
 16 &                          -254
 &                          1703
 &                          9024 \\ 
 17 &                          2524
 &                        -11952
 &                       -152014 \\ 
 18 &                         -6216
 &                         33316
 &                        383616 \\ 
 19 &                          4040
 &                        -18900
 &                       -298186 \\ 
 20 &                         11805
 &                        -64201
 &                       -832320 \\ 
 21 &                        -49400
 &                        304580
 &                       3575922 \\ 
 22 &                         68268
 &                       -401068
 &                      -5486624 \\ 
 23 &                         14928
 &                        -97928
 &                      -1012506 \\ 
 24 &                       -332511
 &                       2390637
 &                      27088992 \\ 
 25 &                        734508
 &                      -5130048
 &                     -65115000 \\ 
 26 &                       -568038
 &                       4264858
 &                      53200524 \\ 
 27 &                      -1641320
 &                      13518716
 &                     147217176 \\ 
 28 &                       6202774
 &                     -49117798
 &                    -608004040 \\ 
 29 &                      -9239676
 &                      76725752
 &                     947874280 \\ 
 30 &                      -2503162
 &                      29308994
 &                     189048900 \\ 
 31 &                      42749908
 &                    -381566684
 &                   -4568526730 \\ 
 32 &                     -99021392
 &                     915306452
 &                   11071969920 \\ 
33 &                      72255812
 &                    -629297848
 &                   -8871938526 \\ 
 34 &                     215763902
 &                   -2149429218
 &                  -24714851124 \\ 
 35 &                    -846523304
 &                    8606730256
 &                  102572776040 \\ 
 36 &                    1235587854
 &                  -12408220218
 &                 -158562077760 \\ 
 37 &                     315695688
 &                   -3956969996
 &                  -31309254516 \\ 
 38 &                   -5897043012
 &                   65853427044
 &                  766255508396 \\ 
 39 &                   13498636700
 &                 -149789004280
 &                -1846277129736 \\ 
\end{tabular}
\newpage
\begin{tabular}{rrrr}
 40 &                  -10063784956
 &                  110599540765
 &                 1479447715520 \\ 
 41 &                  -30197995484
 &                  371951421160
 &                 4133610817968 \\ 
 42 &                  117108185474
 &                -1416033283010
 &               -17054958273276 \\ 
 43 &                 -172710840680
 &                 2102892657652
 &                26339112604404 \\ 
 44 &                  -46214867144
 &                  737547145862
 &                 5331885548880 \\ 
 45 &                  824863285280
 &               -10822599389744
 &              -127080932186700 \\ 
 46 &                -1901022089768
 &                25078129380684
 &               305778947448156 \\ 
 47 &                 1405042568748
 &               -17797597472844
 &              -243733007205368 \\ 
 48 &                 4266178550909
 &               -61005293343300
 &              -684581856372288 \\ 
 49 &               -16624047456088
 &               235876708211784
 &              2818178220557042 \\ 
 50 &                24458757867992
 &              -344426000745528
 &             -4339993392475000 \\ 
 51 &                 6610934151948
 &              -118602900569968
 &              -894117116934894 \\ 
 52 &              -117973371104457
 &              1797119592535141
 &             20963370411907352 \\ 
 53 &               271535984970264
 &             -4116526192115268
 &            -50319881932177670 \\ 
 54 &              -200950868428636
 &              2947822355097388
 &             39972295747477872 \\ 
 55 &              -615007072669600
 &             10130142463339880
 &            112867555200892470 \\ 
 56 &              2391435417419895
 &            -38724154430758393
 &           -463179370952109840 \\ 
 57 &             -3523264660998628
 &             56732375209602912
 &            711889569606231690 \\ 
 58 &              -974049037638220
 &             20114020125177948
 &            149739050620646304 \\ 
 59 &             17078683955539360
 &           -294868317310376404
 &          -3442135262856162448 \\ 
 60 &            -39345145748450867
 &            676479764508534719
 &           8247102824525820120 \\ 
 61 &             29026701896553572
 &           -479158286800083944
 &          -6525890234473448550 \\ 
 62 &             89603802847507184
 &          -1661764311006201920
 &         -18532244655048816588 \\ 
 63 &           -348363066804818696
 &           6354863121022079308
 &          75848843620008360720 \\ 
 64 &            512907395631821606
 &          -9265243259835533768
 &        -116335729831253805824 \\ 
 65 &            144248115519171836
 &          -3315474781302882096
 &         -24961184352327362750 \\ 
 66 &          -2500429353847945250
 &          48350450407929798098
 &         563358836743426377588 \\ 
 67 &           5757911256695782416
 &        -110521765873057849552
 &       -1347235562794556332032 \\ 
 68 &          -4239564000431858236
 &          78112180092393615814
 &        1062223855357818122632 \\ 
 69 &         -13189936451780437660
 &         272577693656067525988
 &        3034102236742714342464 \\ 
 70 &          51212742729615384348
 &       -1038017985499645393024
 &      -12384673817861566133360 \\ 
 71 &         -75378650279043338628
 &        1511592752767037119280
 &       18959244151288425233210 \\ 
 72 &         -21589841096310396846
 &         550095300981147667462
 &        4149995353996807267776 \\ 
 73 &         369164127694023873860
 &       -7896113973546269891772
 &      -91961341446801674710358 \\ 
 74 &        -849854483657640971250
 &       18026239753543948651338
 &      219535041878849931107584 \\ 
 75 &         624038440770346152380
 &      -12669452007330249289520
 &     -172468622446186756857750 \\ 
 76 &        1956508551522393160164
 &      -44535529209867702398308
 &     -495562798199277085255224 \\ 
 77 &       -7587615135291641485816
 &      169246143976718418368880
 &     2017606788248236265104332 \\ 
 78 &       11159761201704504160824
 &     -245838982882938309009072
 &    -3082929124790021245909560 \\ 
 79 &        3251363324100951241776
 &      -90655771470008657225676
 &     -688181679835018200461774 \\
\hline \hline
\end{tabular}
\end{center}
\normalsize 

\hoffset=-1.5cm
\newpage

{\bf \tablespinhalfser:} New low-temperature series for the 
spin-$\frac{1}{2}$ \2D Ising 
magnetisation ($M(u) = \sum_{n}m_{n}u^{n}$), susceptibility
($\chi(u) = \sum_{n}x_{n}u^{n}$), and specific heat
($C_{v}(u) = \sum_{n}c_{n}u^{n}$). All terms with odd $n$ 
are zero.

\footnotesize
\begin{tabular}{rrrr}
\hline \hline
 $n$ & $m_{n}$ & $x_{n}$ & $c_{n}$ \\
\hline
  0 &                               1 & 
                              0 & 
                              0 \\ 
  2 &                               0 & 
                              0 & 
                              0 \\ 
  4 &                              -2 & 
                              1 & 
                             16 \\ 
  6 &                              -8 & 
                              8 & 
                             72 \\ 
  8 &                             -34 & 
                             60 & 
                            288 \\ 
 10 &                            -152 & 
                            416 & 
                           1200 \\ 
 12 &                            -714 & 
                           2791 & 
                           5376 \\ 
 14 &                           -3472 & 
                          18296 & 
                          25480 \\ 
 16 &                          -17318 & 
                         118016 & 
                         125504 \\ 
 18 &                          -88048 & 
                         752008 & 
                         634608 \\ 
 20 &                         -454378 & 
                        4746341 & 
                        3269680 \\ 
 22 &                        -2373048 & 
                       29727472 & 
                       17086168 \\ 
 24 &                       -12515634 & 
                      185016612 & 
                       90282240 \\ 
 26 &                       -66551016 & 
                     1145415208 & 
                      481347152 \\ 
 28 &                      -356345666 & 
                     7059265827 & 
                     2585485504 \\ 
 30 &                     -1919453984 & 
                    43338407712 & 
                    13974825960 \\ 
 32 &                    -10392792766 & 
                   265168691392 & 
                    75941188736 \\ 
 34 &                    -56527200992 & 
                  1617656173824 & 
                   414593263952 \\ 
 36 &                   -308691183938 & 
                  9842665771649 & 
                  2272626444528 \\ 
 38 &                  -1691769619240 & 
                 59748291677832 & 
                 12502223573304 \\ 
 40 &                  -9301374102034 & 
                361933688520940 & 
                 68996534259040 \\ 
 42 &                 -51286672777080 & 
               2188328005246304 & 
                381858968527680 \\ 
 44 &                -283527726282794 & 
              13208464812265559 & 
               2118806030647328 \\ 
 46 &               -1571151822119216 & 
              79600379336505560 & 
              11783826597027256 \\ 
 48 &               -8725364469143718 & 
             479025509574159232 & 
              65674579024955904 \\ 
 50 &              -48552769461088336 & 
            2878946431929191656 & 
             366728645195006000 \\ 
 52 &             -270670485377401738 & 
           17281629934637476365 & 
            2051443799934043632 \\ 
 54 &            -1511484024051198680 & 
          103621922312364296112 & 
           11494250259278105304 \\ 
 56 &            -8453722260102884930 & 
          620682823263814178484 & 
           64499139095733378176 \\ 
 58 &           -47350642314439048648 & 
         3714244852389988540072 & 
          362436080938852037648 \\ 
 60 &          -265579129813183372802 & 
        22206617664989885664363 & 
         2039249170926323834880 \\ 
 62 &         -1491465339550559632448 & 
       132657236460768679560864 & 
        11487673072269872540904 \\ 
 64 &         -8385872784303807639294 & 
       791843294876287279547520 & 
        64786142191741932873984 \\ 
 66 &        -47202746620874986470336 & 
      4723112509660327575046688 & 
       365754067103461706996304 \\ 
 68 &       -265975151780412455885826 & 
     28152514246598001579534217 & 
      2066925549185792626090544 \\ 
 70 &      -1500179080790296495333960 & 
    167696255471026758161692328 & 
     11691314122170272566638200 \\ 
 72 &      -8469330846027919131108866 & 
    998303936498277539688401212 & 
     66188283453887221177721568 \\ 
 74 &     -47856040705247407564621400 & 
   5939502715888619728011515904 & 
    375021938737150106426702208 \\ 
 76 &    -270636033194089067428986890 & 
  35318214476286590871820680287 & 
   2126523853550658555941372768 \\ 
\hline \hline
\end{tabular}
\normalsize 
\hoffset=-1cm

\newpage

{\bf \tablegammauc:} Estimates of $u_{c}$ and $\gamma'$ from 
first and second order differential approximants. $L$ is the order of the
inhomogeneous polynomial.
\begin{center}
\begin{tabular}{lll|ll}
\hline \hline
 & \multicolumn{2}{c|}{First order approximants} 
 & \multicolumn{2}{c}{Second order approximants} \\
\hline
$L$\hspace{1cm} & \multicolumn{1}{c}{$u_{c}$}
& \multicolumn{1}{c|}{$\gamma'$} 
& \multicolumn{1}{c}{$u_{c}$}
& \multicolumn{1}{c}{$\gamma'$} \\ 
\hline
0 & 0.554111(24) &   1.769(6)  & 0.554033(19) & 1.734(9) \\
1 & 0.554105(27) &   1.768(11) & 0.554053(27) & 1.744(11) \\
2 & 0.554076(22) &   1.756(8)  & 0.554057(16) & 1.746(8) \\
3 & 0.554081(13) &   1.756(5)  & 0.554082(28) & 1.757(11) \\
4 & 0.554083(11) &   1.756(5)  & 0.554085(31) & 1.758(12) \\
5 & 0.554078(16) &   1.756(7)  & 0.554071(20) & 1.752(9) \\
6 & 0.554082(8)  &   1.757(3)  & 0.554061(10) & 1.747(5) \\
7 & 0.554079(12) &   1.756(5)  & 0.554061(16) & 1.747(8) \\
8 & 0.554085(19) &   1.759(8)  & 0.554058(15) & 1.745(8) \\ 
\hline \hline
\end{tabular}
\end{center}

\vspace{1cm}

{\bf \tablealphauc:} Estimates of $u_{c}$ and $\alpha'$ from 
first and second order differential approximants. $L$ is the order of the
inhomogeneous polynomial.
\begin{center}
\begin{tabular}{lll|ll}
\hline \hline
 & \multicolumn{2}{c|}{First order approximants} 
 & \multicolumn{2}{c}{Second order approximants} \\
\hline
$L$\hspace{1cm} & \multicolumn{1}{c}{$u_{c}$}
& \multicolumn{1}{c|}{$\alpha'$} 
& \multicolumn{1}{c}{$u_{c}$}
& \multicolumn{1}{c}{$\alpha'$} \\ 
\hline
0  & \multicolumn{1}{c}{------} &  \multicolumn{1}{c|}{------} 
   & 0.554069(41) & 0.001(13) \\
1  & \multicolumn{1}{c}{------} & \multicolumn{1}{c|}{------}
   & 0.554019(33) & 0.019(12) \\
2  & 0.554016(31) &  0.018(11)  & 0.554017(30) & 0.018(12)  \\
3  & 0.554045(31) &  0.009(10)  & 0.554030(33) & 0.015(15) \\
4  & 0.554058(20) &  0.0040(69) & 0.554044(26) & 0.0074(69) \\
5  & 0.554068(25) &  0.0008(88) & 0.554055(32) & 0.0049(98) \\
6  & 0.554053(13) &  0.0062(46) & 0.554061(25) & 0.0030(77) \\
7  & 0.554059(13) &  0.0042(50) & 0.554058(28) & 0.0039(79) \\
8  & 0.554058(14) &  0.0041(48) & 0.554064(24) & 0.0026(72) \\
\hline \hline
\end{tabular}
\end{center}

\newpage

{\bf \tableampl:} Estimates for the critical amplitudes of the 
magnetisation $A_{M}$, the susceptibility $A_{\chi}$ and the 
specific heat $A_{C}$ as obtained from inhomogeneous first order 
differential approximants. $L$ is the order of the inhomogeneous 
polynomial.
\begin{center}
\begin{tabular}{llll}
\hline \hline
$L$\hspace{0.5cm} & \multicolumn{1}{c}{$A_{M}$} & 
\multicolumn{1}{c}{$A_{\chi}$} & 
\multicolumn{1}{c}{$A_{C}$} \\
\hline
4  & 1.2088(18) & 0.0646(58) & 18.98(33) \\
5  & 1.2095(26) & 0.0617(30) & 19.86(69)\\
6  & 1.2090(13) & 0.0629(26) & 20.25(61)\\
7  & 1.2092(12) & 0.0617(20) & 19.95(55)\\
8  & 1.2090(31) & 0.0598(46) & 19.98(65)\\
9  & 1.2093(15) & 0.0599(42) & 19.80(33)\\
10 & 1.2091(5)  & 0.0628(15) & 19.93(31)\\
11 & 1.2089(9)  & 0.0627(14) & 19.78(26)\\
12 & 1.2091(13) & 0.0626(9)  & 19.54(29)\\
13 & 1.2091(5)  & 0.0626(12) & 19.50(44)\\
14 & 1.2089(5)  & 0.0627(11) & 19.56(40)\\
15 & 1.2090(4)  & 0.0625(8)  & 19.54(40)\\
16 & 1.2089(18) & 0.0632(22) & 19.42(44)\\
\hline \hline
\end{tabular}
\end{center}

\vspace{1cm}

{\bf \tablebetadelta:} Estimates for the leading critical exponent
$\beta$ and the confluent exponent $\Delta_{1}$ plus the associated
amplitudes from [N/N+1] Pad\'{e} approximants to the Baker-Hunter 
transformed series for the magnetisation with $u_{c} = 0.554065$.
\begin{center}
\begin{tabular}{ccc|cc}
\hline \hline
N & $\beta$ & $A_{M}$ & $\Delta_{1}$ & $A_{M}a_{M,1}$ \\
\hline
 30  &  0.12480  & 1.2070 &  1.147  & -0.375 \\
 31  &  0.12474  & 1.2066 &  1.180  & -0.424 \\
 32  &  0.12452  & 1.2052 & ------  & ------  \\
 33  &  0.12486  & 1.2074 &  1.120  & -0.343 \\
 34  &  0.12496  & 1.2080 &  1.072  & -0.297 \\
 35  &  0.12495  & 1.2080 &  1.076  & -0.300 \\
 36  &  0.12500  & 1.2084 &  1.050  & -0.280 \\
 37  &  0.12500  & 1.2084 &  1.050  & -0.280 \\
 38  &  0.12498  & 1.2083 &  1.058  & -0.286 \\
 39  &  0.12498  & 1.2082 &  1.063  & -0.290 \\
\hline \hline
\end{tabular}
\end{center}

\newpage

{\bf \tablegammadelta:} Estimates for the leading critical exponent 
$\gamma'$ and the confluent exponent $\Delta_{1}$ plus the associated
amplitudes from [N/N+1] and [N+1/N] Pad\'{e} approximants to the 
Baker-Hunter transformed series for the susceptibility with 
$u_{c} = 0.554065$.
\begin{center}
\begin{tabular}{ccc|cc||cc|cc}
\hline \hline
 & \multicolumn{4}{c||}{[N/N+1]} 
 & \multicolumn{4}{c}{[N+1/N]} \\
\hline
N & $\gamma'$ & $A_{\chi}$ & $\Delta_{1}$ & $A_{\chi}a_{\chi,1}$ &
$\gamma'$ & $A_{\chi}$ & $\Delta_{1}$ & $A_{\chi}a_{\chi,1}$ \\
\hline
 28 & 1.7407 & 0.06475 & 1.235 & 0.5851   
    & 1.7372 & 0.06605 & ------ & ------ \\
 29 & 1.7564 & 0.05991 & 1.110 & 0.4221
    & 1.7481 & 0.06265 & 1.338 & 1.5032 \\
 30 & 1.7529 & 0.06092 & 1.129 & 0.4429
    & 1.7461 & 0.06324 & 1.369 & 1.8334 \\
 31 & 1.7429 & 0.06411 & 1.233 & 0.5908
    & 1.7405 & 0.06502 & ------ & ------ \\
 32 & 1.7494 & 0.06201 & 1.155 & 0.4733
    & 1.7443 & 0.06379 & 1.406 & 2.3719 \\
 33 & 1.7492 & 0.06206 & 1.156 & 0.4750 
    & 1.7441 & 0.06387 & 1.413 & 2.4897 \\
 34 & 1.7498 & 0.06187 & 1.151 & 0.4686
    & 1.7448 & 0.06363 & 1.394 & 2.1671 \\
 35 & 1.7498 & 0.06187 & 1.151 & 0.4684
    & 1.7448 & 0.06362 & 1.393 & 2.1561 \\
 36 & 1.7494 & 0.06199 & 1.155 & 0.4727
    & 1.7448 & 0.06362 & 1.394 & 2.1575 \\
 37 & 1.7490 & 0.06210 & 1.158 & 0.4773
    & 1.7448 & 0.06362 & 1.394 & 2.1633 \\
\hline \hline
\end{tabular}
\end{center}

\vspace{1cm}

{\bf \tabcorr:} The difference between the first correction terms 
with $q=2$ and 3 for the partition function ($\delta_{Z_{0}}$), 
magnetisation ($\delta_{M}$), and susceptibility ($\delta_{\chi}$).
\begin{center}
\begin{tabular}{crrr}
  \hline \hline
  $b_{max}$ & $\delta_{Z_{0}}$ &\hspace{1cm} $\delta_{M} $ &
 \hspace{1cm} $\delta_{\chi}$ \\
  \hline
  2  &  -20    &  180     &    -1620    \\
  3  &  -84    &  1008    &    -12096   \\
  4  &  -220   &  3300    &    -49500   \\
  5  &  -455   &  8190    &    -147420  \\ 
  6  &  -816   &  17136   &    -359856  \\ 
  7  &  -1330  &  31920   &    -766080  \\
  \hline \hline     
\end{tabular} 
\end{center}    

\end{document}